\documentclass[aps,pra,twocolumn,floatfix,superscriptaddress]{revtex4-1}
\usepackage{amsfonts, amssymb, amsmath, graphicx}

\begin{document}

\title{Renormalization of interactions of ultracold atoms in simulated Rashba gauge fields}

\author{Tomoki Ozawa}
\affiliation{
Department of Physics, University of Illinois at Urbana-Champaign, 1110 West Green Street, Urbana, Illinois 61801, USA
}%

\author{Gordon Baym}
\affiliation{
Department of Physics, University of Illinois at Urbana-Champaign, 1110 West Green Street, Urbana, Illinois 61801, USA
}%
\affiliation{
The Niels Bohr International Academy, The Niels Bohr Institute,
Blegdamsvej 17, DK-2100 Copenhagen \O, Denmark
}%

\date{\today}

\def\del{\partial}
\def\p{\prime}
\def\simge{\mathrel{%
         \rlap{\raise 0.511ex \hbox{$>$}}{\lower 0.511ex \hbox{$\sim$}}}}
\def\simle{\mathrel{
         \rlap{\raise 0.511ex \hbox{$<$}}{\lower 0.511ex \hbox{$\sim$}}}}
\newcommand{\feynslash}[1]{{#1\kern-.5em /}}

\begin{abstract}

Interactions of ultracold atoms with Rashba spin-orbit coupling, currently being studied with simulated (artificial) gauge fields, have nontrivial ultraviolet and infrared behavior.   Examining the ultraviolet structure of the Bethe-Salpeter equation, we show that the linear ultraviolet divergence in the bare interaction can be renormalized as usual in terms of low-energy scattering lengths, and that for both bosons and fermions ultraviolet logarithmic divergences are absent.  Calculating the leading order effective interaction with full dependence on the spin-orbit coupling strength and the center-of-mass momentum of the colliding pair, we elucidate the relation between mean-field interactions and physical three-dimensional scattering lengths.   As a consequence of infrared logarithmic divergences in the two-particle propagator, the effective interaction vanishes as the center-of-mass momentum approaches zero.

\end{abstract}

\maketitle

\section{Introduction}

Artificial gauge fields in neutral ultracold atomic systems hold the prospect,  both theoretically and experimentally,
of realizing and exploring a wide variety of new physical systems \cite{Dalibard2010}.
Besides simulated Abelian gauge fields, which give rise to analogs of conventional magnetic fields, artificial
non-Abelian gauge fields, with potential applications to quantum chromodynamics, have been the subject of intense study \cite{Stanescu2008, Ho2010, Wang2010, Zhang, Gopalakrishnan2011, Jian2011, Yip2010, Wu2008, Juzeliunas2010, Campbell2011,Sau2011,Vyasanakere2011a, Vyasanakere2011b,Gong2011,Hu2011,YuZhai,Han2011}.
Certain non-Abelian gauge fields are equivalent to Rashba-Dresselhaus spin-orbit interactions \cite{Rashba1960, Dresselhaus1955}; 
the first experimental realization of such a spin-orbit coupled ultracold system was reported in Ref.~\cite{Lin2011}.
Striped phases and topologically non-trivial states have been predicted, within mean-field theory, in atomic bosonic systems with a Rashba-type spin-orbit coupling \cite{Wang2010,Jian2011}.

A Rashba spin-orbit interaction, $\kappa \mathbf{p}_\perp \cdot \boldsymbol\sigma$, with $\kappa$ the Rashba coupling and $\mathbf{p}_\perp = (p_x, p_y, 0)$, leads to two non-quadratic branches in the single particle dispersion relation [Eq.~(\ref{dispersion})].
As a consequence, the effective interactions between particles can exhibit nontrivial ultraviolet and infrared structure.  This structure is in contrast to usual dilute three-dimensional ultracold atomic systems where mean-field interactions are proportional to three-dimensional scattering lengths.
Our purpose in this paper is to elucidate the ultraviolet and infrared structure of the effective interactions in ultracold atomic systems with Rashba spin-orbit coupling, and their relation to physically observable scattering lengths, which is needed to analyze such systems both theoretically and experimentally.
 
The ultraviolet and infrared problems have been considered earlier.  Yang and Sachdev \cite{Yang2006} analyzed a condensed matter system with one branch with a dispersion relation similar to the lower energy branch with a Rashba interaction, $\epsilon (\mathbf{p}) = [(p - \kappa)^2]/2m$, obtaining both ultraviolet and infrared cutoff-dependent logarithmic contributions to the effective interaction.
Similarly, Gopalakrishnan {\it et al}. \cite{Gopalakrishnan2011} have recently considered one-loop corrections to the effective interaction in atomic systems within the lower energy branch of a Rashba interaction and also obtained a logarithmic dependence on the ultraviolet and infrared cutoffs.
The arguments in Refs.~\cite{Yang2006} and \cite{Gopalakrishnan2011} are mainly in two dimensions, but a similar argument applied to three dimensions with one branch of dispersion leads to the logarithmic dependence on the ultraviolet cutoff as well.
The essential argument is that with a single branch, the two particle Green's function in the t-matrix equation has the ultraviolet behavior 
\begin{align}
    \int^\Lambda \frac{d^3k}{(2\pi)^3} \frac{m}{k_z^2 + (|k_\perp| - \kappa)^2} \to  
     \frac{m}{2\pi^2}\left( \Lambda + \frac{\pi\kappa}{2}\ln\Lambda +\cdots\right),
\end{align}
where $\Lambda$ is the ultraviolet momentum cutoff.
We show here that with the doubly branched dispersion taken into account, logarithmic ultraviolet divergences, in fact, are absent in three dimensions so that
the linear ultraviolet divergence in two-particle scattering can be nicely renormalized away as usual, and the effective interaction depends solely on low-energy scattering lengths.
Similarly, logarithmic ultraviolet divergences do not appear in the gap equation for fermions in paired states, and linear divergences can again be renormalized away in favor of scattering lengths, as discussed in Refs.~\cite{Vyasanakere2011a, Vyasanakere2011b,Gong2011, Hu2011,YuZhai,Han2011,ZhaiUnpublished}.

Even though after renormalization the effective interaction becomes free of ultraviolet divergences,
the full relation of the effective interactions in systems with Rashba spin-orbit coupling to physically observable scattering lengths has interesting structure, owing in part to an infrared logarithmic divergence in the  two particle propagator
entering the t-matrix when the pair's center-of-mass momentum, $\mathbf{q}$, goes to zero. In essence,
\begin{align}
	\int_0 \frac{d^3 k}{(2\pi)^3}
	&\frac{2m}{(|\mathbf{k}_\perp+\mathbf{q}/2| - \kappa)^2 + 
	(|- \mathbf{k}_\perp+\mathbf{q}/2 | - \kappa)^2 + 2k_z^2
	} 
	\notag \\
	&\to
	-\frac{m\kappa}{2\pi}\ln q. \label{infrared}
\end{align}
As a consequence, the effective two-particle interaction vanishes as $q\to 0$.  This infrared divergence, which can be obtained within a single branch model, as in Refs.~\cite{Yang2006,Gopalakrishnan2011}, is a result of an infinity of pairs of zero energy states with zero center-of-mass momentum.   We construct here the full solution of the two-particle Bethe-Salpeter equation for the t-matrix for fermions, and an approximate t-matrix for bosons valid when the free space three-dimensional scattering lengths are small, including exactly the two branches of the single-particle dispersion relation and the infrared structure.

\section{Particles with Rashba interaction}

We consider both bosons and fermions in three dimensions with two internal states $a$ and $b$,  and with a symmetric Rashba spin-orbit interaction described by the Hamiltonian,
\begin{align}
	\mathcal{H}
	&=
	\sum_{\mathbf{p}}
	\begin{pmatrix}
	a_\mathbf{p}^\dagger & b_\mathbf{p}^\dagger
	\end{pmatrix}
	\left[
	\frac{p^2 + \kappa^2}{2m}I + \frac{\kappa}{m} (\sigma_x p_x + \sigma_y p_y)
	\right]
	\begin{pmatrix}
	a_\mathbf{p} \\
	b_\mathbf{p}
	\end{pmatrix}
	\notag \\
	&+
	\frac{1}{2V}
	\sum_{\mathbf{p}_1 + \mathbf{p}_2 = \mathbf{p}_3 + \mathbf{p}_4}
	\left(
	g_{aa} a_{\mathbf{p}_4}^\dagger a_{\mathbf{p}_3}^\dagger a_{\mathbf{p}_2} a_{\mathbf{p}_1}
	\right.
	\notag \\
	&\hspace{1cm}\left.
	+
	g_{bb} b_{\mathbf{p}_4}^\dagger b_{\mathbf{p}_3}^\dagger b_{\mathbf{p}_2} b_{\mathbf{p}_1}
	+
	2g_{ab}\ a_{\mathbf{p}_4}^\dagger b_{\mathbf{p}_3}^\dagger b_{\mathbf{p}_2} a_{\mathbf{p}_1}
	\right).
\end{align}
Here, $m$ is the atomic mass, $V$ is the volume of the system, $a_\mathbf{p}$ annihilates an atom in internal state $a$ with momentum $\mathbf{p}$, and $b_\mathbf{p}$ annihilates an atom in state $b$ with momentum $\mathbf{p}$; the $\sigma_x$ and $\sigma_y$ are the usual Pauli matrices between the internal states, and 
$I$ is the two-by-two identity matrix.  We take the coupling $\kappa$ to be positive.  The $g_{aa}$, $g_{bb}$, and $g_{ab}$ are the bare s-wave couplings between $a$-$a$, $b$-$b$, and $a$-$b$ particles; for fermions, $g_{aa}$, $g_{bb}$ are not present.

The single particle part of the Hamiltonian can be diagonalized by introducing operators $\alpha_\mathbf{p}$ and $\beta_\mathbf{p}$ by
\begin{align}
	\begin{pmatrix}
	\alpha_\mathbf{p} \\
	\beta_\mathbf{p}
	\end{pmatrix}
	=
	\frac{1}{\sqrt{2}}
	\begin{pmatrix}
	1 & -e^{-i\phi} \\
	1 & e^{-i\phi}
	\end{pmatrix}
	\begin{pmatrix}
	a_\mathbf{p} \\
	b_\mathbf{p}
	\end{pmatrix},
\end{align}
where $\phi$ is the angle of $\mathbf{p}$ in the $x$-$y$ plane;
the Hamiltonian becomes
\begin{align}
	\mathcal{H}
	&=
	\sum_\mathbf{p} \left( \epsilon_- (\mathbf{p}) \alpha^\dagger_\mathbf{p} \alpha_\mathbf{p} + \epsilon_+ (\mathbf{p}) \beta^\dagger_\mathbf{p} \beta_\mathbf{p} \right)
	+\mathcal{H}_{\mathrm{int}}.
\end{align}
The single-particle spectrum has two branches,
\begin{align}
	\epsilon_\pm (\mathbf{p}) = \frac{(p_\perp \pm \kappa)^2 + p_z^2}{2m},
	\label{dispersion}
\end{align}
with the lower branch $\epsilon_- (\mathbf{p})$ having degenerate single-particle ground states on the circle $p_\perp = \kappa$.

The interaction part of the Hamiltonian, $\mathcal{H}_{\mathrm{int}}$, in the  $\alpha$-$\beta$ basis becomes
\begin{align}
	&\mathcal{H}_{\mathrm{int}}
	=
	\frac{1}{V}
	\sum_{\mathbf{p}_1 + \mathbf{p}_2 = \mathbf{p}_3 + \mathbf{p}_4}
	\notag \\
	&
	\left[
	\mathcal{V}^{(1)}_{\phi_1, \phi_2; \phi_3, \phi_4}
	\left(
	\alpha_{\mathbf{p}_4}^\dagger \alpha_{\mathbf{p}_3}^\dagger \alpha_{\mathbf{p}_2} \alpha_{\mathbf{p}_1}
	+
	\beta_{\mathbf{p}_4}^\dagger \beta_{\mathbf{p}_3}^\dagger \beta_{\mathbf{p}_2} \beta_{\mathbf{p}_1}
	\right)/2
	\right.
	\notag \\
	&
	+
	\mathcal{V}^{(2)}_{\phi_1, \phi_2; \phi_3, \phi_4}
	\left(
	\beta_{\mathbf{p}_4}^\dagger \beta_{\mathbf{p}_3}^\dagger \alpha_{\mathbf{p}_2} \alpha_{\mathbf{p}_1}
	+
	\alpha_{\mathbf{p}_4}^\dagger \alpha_{\mathbf{p}_3}^\dagger \beta_{\mathbf{p}_2} \beta_{\mathbf{p}_1}
	\right)/2
	\notag\\
	&
	+
	\mathcal{V}^{(3)}_{\phi_1, \phi_2; \phi_3, \phi_4}
	\left(
	\alpha_{\mathbf{p}_4}^\dagger \beta_{\mathbf{p}_3}^\dagger \beta_{\mathbf{p}_2} \beta_{\mathbf{p}_1}
	+
	\beta_{\mathbf{p}_4}^\dagger \alpha_{\mathbf{p}_3}^\dagger \alpha_{\mathbf{p}_2} \alpha_{\mathbf{p}_1}
	\right)/\sqrt{2}
	\notag \\
	&
	+\mathcal{V}^{(4)}_{\phi_1, \phi_2; \phi_3, \phi_4}
	\left(
	\alpha_{\mathbf{p}_4}^\dagger \alpha_{\mathbf{p}_3}^\dagger \beta_{\mathbf{p}_2} \alpha_{\mathbf{p}_1}
	+
	\beta_{\mathbf{p}_4}^\dagger \beta_{\mathbf{p}_3}^\dagger \alpha_{\mathbf{p}_2} \beta_{\mathbf{p}_1}
	\right)/\sqrt{2}
	\notag \\
	&\left.+
	\mathcal{V}^{(5)}_{\phi_1, \phi_2; \phi_3, \phi_4}
	\alpha_{\mathbf{p}_4}^\dagger \beta_{\mathbf{p}_3}^\dagger \beta_{\mathbf{p}_2} \alpha_{\mathbf{p}_1}
	\right], \label{hamalphabeta}
\end{align}
where $\phi_i$ is the angle of $\mathbf{p}_i$ in the $x$-$y$ plane, and
where
\begin{align}
	\mathcal{V}^{(1)}_{\phi_1, \phi_2; \phi_3, \phi_4}
	&=
	A_+ \pm
	\frac{g_{ab}}{8}\left( e^{i\phi_1} \pm e^{i\phi_2} \right) \left( e^{-i\phi_3} \pm e^{-i\phi_4} \right)
	\notag\\
	\mathcal{V}^{(2)}_{\phi_1, \phi_2; \phi_3, \phi_4}
	&=A_+ \mp
	\frac{g_{ab}}{8}\left( e^{i\phi_1} \pm e^{i\phi_2} \right) \left( e^{-i\phi_3} \pm e^{-i\phi_4} \right)
	\notag\\
	\mathcal{V}^{(3)}_{\phi_1, \phi_2; \phi_3, \phi_4}
	&=
	\sqrt{2}A_-\pm
	\frac{g_{ab}}{4\sqrt{2}}\left( e^{i\phi_1} \pm e^{i\phi_2} \right) \left( e^{-i\phi_3} \mp e^{-i\phi_4} \right)
	\notag\\
	\mathcal{V}^{(4)}_{\phi_1, \phi_2; \phi_3, \phi_4}
	&=
	\sqrt{2}A_-\pm
	\frac{g_{ab}}{4\sqrt{2}}\left( e^{i\phi_1} \mp e^{i\phi_2} \right) \left( e^{-i\phi_3} \pm e^{-i\phi_4} \right)
	\notag\\
	\mathcal{V}^{(5)}_{\phi_1, \phi_2; \phi_3, \phi_4}
	&= 2A_+
	\mp
	\frac{g_{ab}}{4}\left( e^{i\phi_1} \mp e^{i\phi_2} \right) \left( e^{-i\phi_3} \mp e^{-i\phi_4} \right), \label{vertices}
\end{align}
with
\begin{align}
  A_\pm &= \left(g_{aa} \pm g_{bb} e^{i(\phi_1+\phi_2-\phi_3-\phi_4)}\right)/4,
\end{align}
where the upper signs are for bosons and the lower signs are for fermions in Eq. (\ref{vertices}).
The denominators $\sqrt{2}$ and $2$ in Eq. (\ref{hamalphabeta}) are chosen so that the vertices in the Feynman rules are just the $\mathcal{V}^{(i)}$s.

\section{t-matrix}

\begin{figure}[htbp]
\begin{center}
\includegraphics[width=8cm]{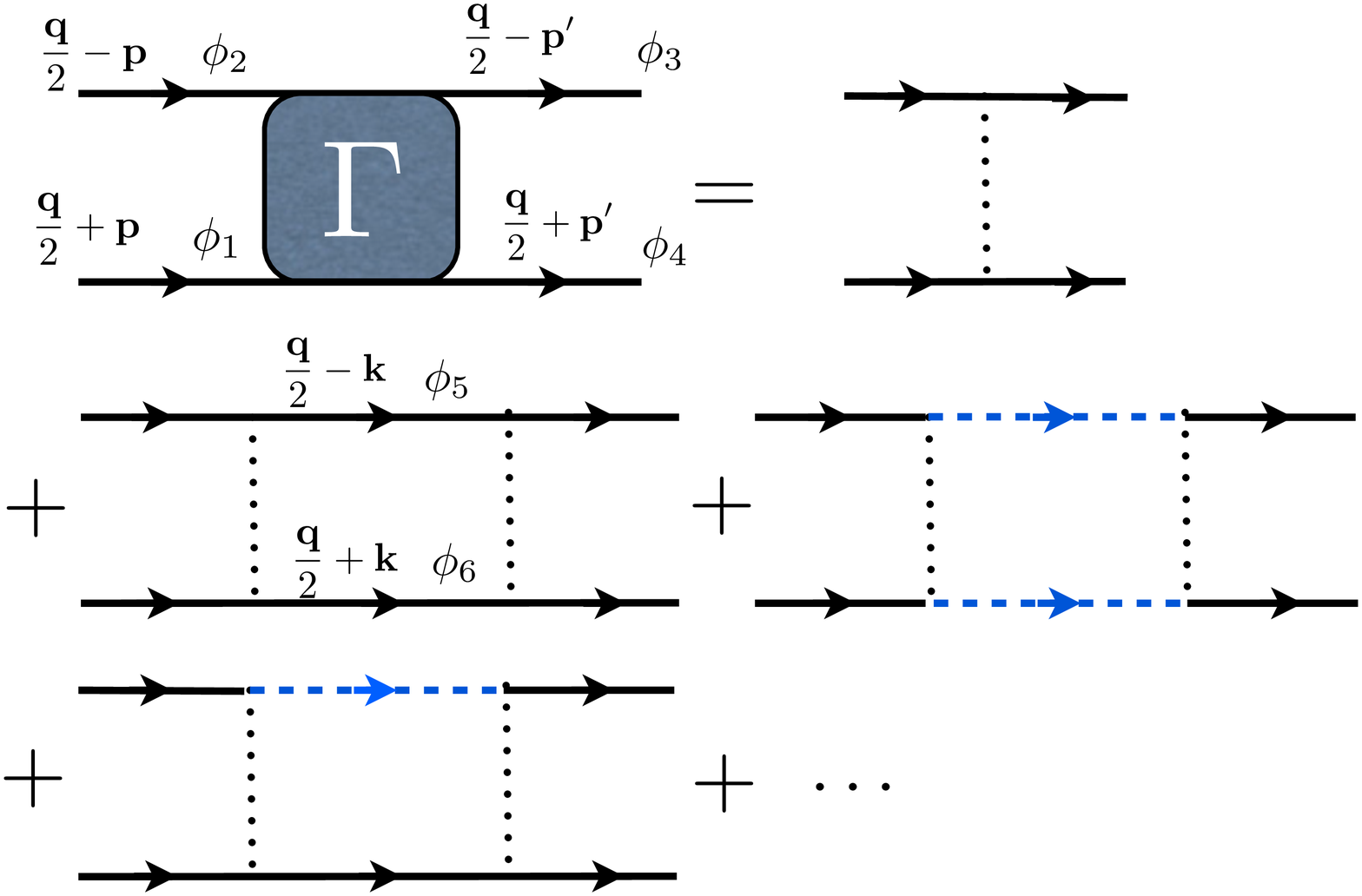}
\caption{The scattering t-matrix for two particles in the $\alpha$-branch.  The solid lines denote particles in $\alpha$-branch, and the dashed lines are particles in the $\beta$-branch. The $\phi_i$ are the angles of the corresponding momenta in the $x$-$y$ plane.}
\label{secondorder_aaaa}
\end{center}
\end{figure}

To obtain the low-energy effective interaction, we construct the t-matrix describing the collision of two atoms in the $\alpha$-branch with incoming momenta $\mathbf{q}/2 + \mathbf{p}$ and $\mathbf{q}/2 - \mathbf{p}$ and outgoing momenta $\mathbf{q}/2 + \mathbf{p}^\prime$ and $\mathbf{q}/2 - \mathbf{p}^\prime$.
The momentum of each particle is on the degenerate ground-state circle.  The single particle propagators are $1/(\omega - \epsilon_\pm (\mathbf{p}))$, and characteristically,  the interactions in the $\alpha$-$\beta$ basis are dependent on angle. 
The t-matrix is the sum of ladder diagrams (Fig. \ref{secondorder_aaaa}).
We denote the momenta of particles in the intermediate off-shell states by $\mathbf{q}/2 + \mathbf{k}$ and $\mathbf{q}/2 - \mathbf{k}$ and label angles  of the momenta $\phi_i$ as in the figure.
The Bethe-Salpeter equation for the zero-energy vertex function $\Gamma_{\alpha \alpha}^{\alpha \alpha}$ with incoming and outgoing $\alpha$-$\alpha$ particles is
\begin{align}
	&\Gamma_{\alpha \alpha}^{\alpha \alpha} (\mathbf{p}, \mathbf{p}^\prime ; \mathbf{q})
	=
	\mathcal{V}^{(1)}_{\phi_1, \phi_2; \phi_3, \phi_4}
	\notag\\
	&
	-
	\int \frac{d^3 k}{(2\pi)^3}
	\left[
	\frac{\mathcal{V}^{(1)}_{\phi_1, \phi_2; \phi_5, \phi_6} \Gamma_{\alpha \alpha}^{\alpha \alpha}(\mathbf{k}, \mathbf{p}^\prime; \mathbf{q})}{\epsilon_- (\frac{\mathbf{q}}{2} - \mathbf{k}) + \epsilon_- (\frac{\mathbf{q}}{2}+\mathbf{k})} \right. 
\notag\\
 &+
	\frac{\mathcal{V}^{(2)}_{\phi_1, \phi_2; \phi_5, \phi_6} \Gamma_{\beta \beta}^{\alpha \alpha}(\mathbf{k}, \mathbf{p}^\prime; \mathbf{q})}{\epsilon_+ (\frac{\mathbf{q}}{2} - \mathbf{k}) + \epsilon_+ (\frac{\mathbf{q}}{2}+\mathbf{k})}
	\pm
\frac{\mathcal{V}^{(3)}_{\phi_1, \phi_2; \phi_6, \phi_5} \Gamma_{\alpha \beta}^{\alpha \alpha}(\mathbf{k},\mathbf{p}^\prime; \mathbf{q})}{\epsilon_+ (\frac{\mathbf{q}}{2} - \mathbf{k}) + \epsilon_- (\frac{\mathbf{q}}{2}+\mathbf{k})}
\left.	\right], \label{bethesalpeteraaaa}
\end{align}
where $\Gamma_{\beta \beta}^{\alpha \alpha}$ is the t-matrix for incoming $\beta$-$\beta$ particles colliding and becoming $\alpha$-$\alpha$ particles, $\Gamma_{\alpha \beta}^{\alpha \alpha}$ is the t-matrix for incoming $\alpha$-$\beta$ colliding and becoming $\alpha$-$\alpha$ particles, and 5 and 6 label the intermediate states. The $\Gamma_{\beta \beta}^{\alpha \alpha}$ and $\Gamma_{\alpha \beta}^{\alpha \alpha}$ obey similar Bethe-Salpeter equations.
We note that $\mathbf{p}$ and $\mathbf{p}^\prime$ dependences of $\Gamma (\mathbf{p}, \mathbf{p}^\prime; \mathbf{q})$ are only through the angles $\phi_i$.
Counting powers in a term-by-term expansion of the t-matrix, one finds that the kernels are linearly divergent  in the ultraviolet in each  order in the expansion, since the denominators have two powers of $k$ while the integral is over three dimensions.

We first show that there is no ultraviolet logarithmic divergence in the t-matrix.  For fermions the Bethe-Salpeter equation for $\Gamma_{\beta \beta}^{\alpha \alpha}(\mathbf{p}, \mathbf{p}^\prime ; \mathbf{q})$ implies that $\Gamma_{\alpha \alpha}^{\alpha \alpha}(\mathbf{p}, \mathbf{p}^\prime ; \mathbf{q}) = -\Gamma_{\beta \beta}^{\alpha \alpha}(\mathbf{p}, \mathbf{p}^\prime ; \mathbf{q})$.
Also, since $\mathcal{V}^{(1)}_{\phi_1, \phi_2; \phi_5, \phi_6} = -\mathcal{V}^{(2)}_{\phi_1, \phi_2; \phi_5, \phi_6}$, the numerators of the first two terms in Eq.~(\ref{bethesalpeteraaaa}) are equal, and depend on $\mathbf{k}$ only through the angles of $\mathbf{q}/2\pm\mathbf{k}$. 
The first two terms when combined have a linear but no logarithmic divergence.
The denominator in the third term is explicitly a function of $k_\perp^2$ and $k_z^2$ for large $k$ and, thus, the integral also does not contain an ultraviolet logarithmic divergence.  The entire integral in Eq.~(\ref{bethesalpeteraaaa}) has only a linear divergence.

For bosons, we note that for large  $k$,
$\phi_6 \sim \phi_k$ and $\phi_5 \sim \phi_k + \pi$, where $\phi_k$ is the angle of $\mathbf{k}$ in the $x$-$y$ plane; hence, $e^{i\phi_5} + e^{i\phi_6} =  \mathcal{O}(k^{-1})$, and $\mathcal{V}^{(1)}_{\phi_1, \phi_2; \phi_5, \phi_6} = \mathcal{V}^{(2)}_{\phi_1, \phi_2; \phi_5, \phi_6}$ to leading order.
Also, the Bethe-Salpeter equations for $\Gamma_{\alpha \alpha}^{\alpha \alpha}(\mathbf{k}, \mathbf{p}^\prime; \mathbf{q})$ and $\Gamma_{\beta \beta}^{\alpha \alpha}(\mathbf{k}, \mathbf{p}^\prime; \mathbf{q})$ are the same to leading order when $k$ is large. Thus, the same mechanism as for fermions leads to a cancellation of the logarithmic divergences, leaving only with the linear divergence, on which we now focus.

Looking at the Bethe-Salpeter equation as a perturbation series, we see that in the intermediate processes, to leading order in $1/k^2$ the coupling $g_{bb}$ appears in the vertices multiplied by a factor of $e^{2i\phi_k}$ and the coupling $g_{ab}$ by $e^{i\phi_k}$, while the coupling $g_{aa}$ is not multiplied by a phase factor.
This implies that in integration over intermediate momenta, cross terms between different couplings do not lead to linear divergences due to the phase integrals, whereas the phase factors in terms with the same couplings are canceled and form geometric series.

\begin{figure}[htbp]
\begin{center}
\includegraphics[width=8cm]{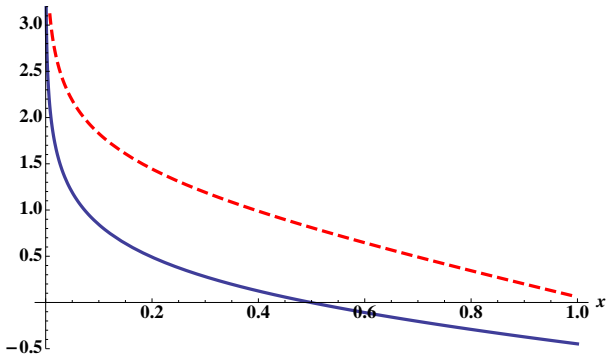}
\caption{The center-of-mass momentum dependence of $f(x)$ (solid line) and $g(x)$ (dashed line) entering the renormalized interaction [Eq.~(\ref{effint})].}
\label{fxgx}
\end{center}
\end{figure}

For bosons, we obtain an approximate effective interaction by adding linearly divergent terms,
which yield the leading-order terms in the physical scattering lengths, $a_{ij}$, times $\kappa$.
The higher-order terms for bosons include cross terms between different scattering lengths,
which do not suffer from ultraviolet divergences.
The linear divergences sum to give the effective boson interaction:
\begin{align}
	\Gamma_{\alpha \alpha}^{\alpha \alpha}&(\mathbf{p}, \mathbf{p}^\prime ; \mathbf{q})
	\notag \\
	\sim&
	\frac{\pi a_{aa}}{m} \frac{1}{1+ a_{aa} \kappa f(\tilde{q}/2)}
	+
	\frac{\pi a_{bb}}{m} \frac{e^{i (\phi_1 + \phi_2 - \phi_3 - \phi_4)}}{1+ a_{bb} \kappa f(\tilde{q}/2)} 
	\notag \\
	& +
	\frac{\pi a_{ab}}{2m} \frac{(e^{i\phi_1} + e^{i\phi_2}) (e^{-i\phi_3} + e^{-i\phi_4})}{1+ a_{ab} \kappa (f(\tilde{q}/2) - g(\tilde{q}/2))},
	\label{effint}
\end{align}
where 
\begin{align}
	 \frac{m}{4\pi a_{ij}} &=  \frac{1}{g_{ij}}+ \frac{m\Lambda}{2\pi^2}
\end{align}
defines the three physical three-dimensional scattering lengths, $a_{aa}$, $a_{bb}$, and $a_{ab}$, in the absence of spin-orbit coupling in terms of the bare couplings and the ultraviolet cutoff $\Lambda$.
The functions $f(\tilde{q}/2)$ and $g(\tilde{q}/2)$ are 
\begin{align}
	&f(\tilde{q}/2)
	\equiv
	\frac{\pi}{m \kappa}\int \frac{d^3 k}{(2\pi)^3}
	\left[
	\frac{1}{\epsilon_- (\frac{\mathbf{q}}{2} + \mathbf{k}) + \epsilon_- (\frac{\mathbf{q}}{2} - \mathbf{k})}
	+
	\right.
	\notag \\
	&
	\left.
	\frac{1}{\epsilon_+ (\frac{\mathbf{q}}{2} + \mathbf{k}) + \epsilon_+ (\frac{\mathbf{q}}{2} - \mathbf{k})}
	+\frac{2}{\epsilon_- (\frac{\mathbf{q}}{2} + \mathbf{k}) + \epsilon_+ (\frac{\mathbf{q}}{2} - \mathbf{k})}
	-
	\frac{4m}{k^2}
	\right]  \notag \\
	&g(\tilde{q}/2)
	\equiv
	-\frac{\pi}{m \kappa} \int \frac{d^3 k}{(2\pi)^3}
	\left[
	\frac{\cos (\phi_5 - \phi_6)}{\epsilon_- (\frac{\mathbf{q}}{2} + \mathbf{k}) + \epsilon_- (\frac{\mathbf{q}}{2} - \mathbf{k})}
	\right.
	\notag \\
	&
	+\left.
	\frac{\cos (\phi_5 - \phi_6)}{\epsilon_+ (\frac{\mathbf{q}}{2} + \mathbf{k}) + \epsilon_+ (\frac{\mathbf{q}}{2} - \mathbf{k})}
	-\frac{2\cos (\phi_5 - \phi_6)}{\epsilon_- (\frac{\mathbf{q}}{2} + \mathbf{k}) + \epsilon_+ (\frac{\mathbf{q}}{2} - \mathbf{k})}
	\right],
\end{align}
where $\tilde{q} \equiv q/\kappa$.
When the colliding particles have zero energy, the center of mass momentum $q$ can be at most $2\kappa$,
which gives the restriction $0 \le \tilde{q}/2 \le 1$.
The functions $f(\tilde{q}/2)$ and $g(\tilde{q}/2)$, which do not diverge in the ultraviolet, are plotted in Fig. \ref{fxgx}.
The effective interaction Eq.~(\ref{effint}) is free of ultraviolet divergence.

 For fermions, where $a_{aa} = a_{bb} = 0$, no cross terms exist and the geometric series obtained by adding linearly divergent terms is an exact solution of the Bethe-Salpeter equation:
 \begin{align}
	\Gamma_{\alpha \alpha}^{\alpha \alpha}&(\mathbf{p}, \mathbf{p}^\prime ; \mathbf{q})
	=-
	\frac{\pi a_{ab}}{2m} \frac{(e^{i\phi_1} - e^{i\phi_2}) (e^{-i\phi_3} - e^{-i\phi_4})}{1+ a_{ab} \kappa (f(\tilde{q}/2) + g(\tilde{q}/2))}.
	\label{effintferm}
\end{align}

Both functions $f(\tilde{q}/2)$ and $g(\tilde{q}/2)$ diverge logarithmically at $q = 0$, as noted earlier in Eq. (\ref{infrared}), leading to the logarithmic vanishing of the effective interaction as $q \to 0$, in agreement with the result obtained by Gopalakrishnan {\it et al}. \cite{Gopalakrishnan2011}.
This infrared divergence arises from the existence of infinitely many pairs of zero-energy single-particle states with $q = 0$, which add up to give an infrared divergence in the two-particle propagator in the t-matrix. For $\mathbf{q} \neq 0$, there is only one pair of zero-energy states and, thus, no infrared divergence.

The result derived here implies that, when the spin-orbit coupling is small ($a_{ij} \kappa \ll 1$), the effective interaction at
$q$ away from 0 is essentially given by replacing the bare couplings by scattering lengths:  
$g_{ij} \to 4\pi a_{ij}/m$, but 
around $q\sim 0$, the effective interaction essentially depends on the center-of-mass momentum and vanishes with an inverse logarithm as $q \to 0$.
For stronger spin-orbit coupling (larger $a_{ij}\kappa < 1$), the center-of-mass dependence of the effective interaction comes into play for a large range of $\tilde{q}$.
For bosons with even stronger spin-orbit coupling ($a_{ij} \kappa \simge 1$),  we need to include higher-order terms in the t-matrix  beyond those included in Eq.~(\ref{effint}) to obtain the effective interaction.

The gap equation for fermion pairing of $a$-fermions with $b$-fermions in a singlet channel has a structure similar to the first two terms in Eq.~(\ref{bethesalpeteraaaa}), which, with the two terms combined, contain only a linear and not a logarithmic divergence 
 \cite{Vyasanakere2011a, Vyasanakere2011b,Gong2011,Hu2011,YuZhai,Han2011,ZhaiUnpublished}. The linear divergence can be renormalized away in favor of  the three-dimensional scattering length, a
 procedure requiring the existence of two dispersion branches.  Studies with only one branch of the dispersion relation yields, as noted, logarithmic ultraviolet divergences in the effective interaction\cite{Yang2006,Gopalakrishnan2011}.

The implications of the effective interaction derived here for the ground state and the finite temperature properties of many-body Bose and Fermi systems will be discussed in a future publication.

\begin{acknowledgements}
This research was supported in part by NSF Grant Nos. PHY07-0701611 and PHY09-69790.  Author T.O. thanks Sarang Gopalakrishnan for stimulating discussions, and author G.B. is grateful to Hui Zhai, Xiaoling Cui, and Yusuke Nishida, for sharing their insights into this problem, and to the Aspen Center for Physics where parts of this research were carried out. 
\end{acknowledgements}

\end{document}